\documentclass[conference]{IEEEtran}
\usepackage{times}

\usepackage[numbers]{natbib}
\usepackage[bookmarks=true]{hyperref}
\usepackage{booktabs} 

\usepackage{graphicx,caption,subcaption}
\usepackage{amsmath,amssymb,stmaryrd,mathtools}
\usepackage{flushend}
\usepackage[ruled,vlined,titlenotnumbered,linesnumbered]{algorithm2e} 
\usepackage[noend]{algpseudocode}
\usepackage{acronym}
\usepackage{comment}
\usepackage{color}
\usepackage{units}

\usepackage[capitalise]{cleveref}

\DeclareCaptionLabelSeparator{periodspace}{.\quad}
\crefname{equation}{}{} 
\crefname{section}{Sec.}{Sec.}

\newcommand{\sys}{\mathcal{S}}
\newcommand{\abs}{\mathcal{M}}
\newcommand{\taskspace}{\mathcal{G}}
\newcommand{\task}{g}
\newcommand{\dist}{d_{\taskspace}}
\newcommand{\paramspace}{\mathcal{P}}
\newcommand{\param}{p}

\newcommand{\state}{x}
\newcommand{\statespace}{\mathcal{X}}
\newcommand{\controlspace}{\mathcal{U}}
\newcommand{\traj}{\xi}

\newcommand{\ctrl}{u}
\newcommand{\cont}{\pi}
\renewcommand{\time}{t}
\newcommand{\cost}{J}
\newcommand{\horizon}{H}
\newcommand{\stateseq}{\bold{\state}}
\newcommand{\ctrlseq}{\bold{\ctrl}}
\newcommand{\totalcost}{\cost(\stateseq, \ctrlseq)}

\newcommand{\dataset}[0]{\mathcal{D}}

\newcommand{\mat}[1]{\boldsymbol{\uppercase{#1}}}
\renewcommand{\vec}[1]{\boldsymbol{#1}}

\newcommand{\acqfuncNo}[0]{\alpha}				
\newcommand{\acqfunc}[1]{\acqfuncNo \left( #1 \right)}		

\newcommand{\reals}{\mathbb{R}}
\newcommand{\argmax}{\operatornamewithlimits{argmax}}

\newcommand{\metName}{CuSTOM}

\newtheorem{remark}{Remark}
\newtheorem{definition}{Definition}

\newtheorem{corollary}{Corollary}
\newtheorem{proposition}{Proposition}
\newtheorem{example}{Example}

\IEEEoverridecommandlockouts

\begin{document}
\title{\LARGE \bf Context-Specific Validation of Data-Driven Models} 
\author{Somil Bansal*, Shromona Ghosh*, Alberto Sangiovanni-Vincentelli, Sanjit A. Seshia, Claire J. Tomlin 
\thanks{* These authors contributed equally to this work. All authors are with the Department of Electrical Engineering and Computer Sciences, University of California, Berkeley. \{somil, shromona.ghosh, alberto, sseshia, tomlin\}@eecs.berkeley.edu.}
}
\maketitle

\begin{abstract}
With an increasing use of data-driven models to control robotic systems, it has become important to develop a methodology for validating such models before they can be deployed to design a controller for the actual system.
Specifically, it must be ensured that the controller designed for a learned model would perform as expected on the actual physical system. 
We propose a context-specific validation framework to quantify the quality of a learned model based on a distance measure between the closed-loop actual system and the learned model.
We then propose an active sampling scheme to compute a probabilistic upper bound on this distance in a sample-efficient manner.
The proposed framework validates the learned model against only those behaviors of the system that are relevant for the purpose for which we intend to use this model, and does not require any \textit{a priori} knowledge of the system dynamics.
Several simulations illustrate the practicality of the proposed framework for validating the models of real-world systems, and consequently, for controller synthesis.
\end{abstract}


%
\section{Introduction} \label{sec:intro}
%
Recently, research in robotics and control theory has been focusing on developing complex autonomous systems, such as robotic manipulators, autonomous cars, surgical robots, etc.
Synthesizing controllers for such complex systems is a challenging but an important problem.
Moreover, we need to ensure that the synthesized controller behaves as intended on the \textit{actual} system. 
This is particularly important for safety-critical systems, where the actions of an autonomous agent may affect human lives. 
This motivates us to develop formal techniques to reason about the performance of synthesized controllers, before deploying them in the real world.

This is often done by identifying an open-loop model and comparing it with the actual system; a controller is then designed on the model, which is tried out in simulation, and subsequently, tested on the actual system.
Although the models obtained and tested in this fashion are more general and robust, this scheme can be prohibitive for complex autonomous systems, especially when the functional form of the dynamics is unknown.
Moreover, we often care about controlling the system on a specific set of tasks, where a coarse dynamics model is often sufficient for control purposes.
In such cases, an abstraction (or model) of a system is obtained and formally \textit{validated} before being used for synthesizing controllers.
Recently, there has been an increased interest in using machine learning (ML) and AI for learning an abstraction \textit{directly} based on the data collected on the system~\cite{bansal2016learning, Bansal2017a}. 
Thus, it has become even more important to develop validation schemes for such abstractions, and is the main focus of this paper.

Broadly, validation refers to formally quantifying the quality/ability of a model to mimic the (dynamic) behavior of the actual system.
Several methods have been proposed in literature to validate an abstraction (or model), $\abs$, of the actual system, $\sys$, utilizing the notion of \textit{simulation metric}, where the quality of an abstraction is quantified via metrics that specify how close the trajectories of $\sys$ and $\abs$ are.
This approach has led to the study of exact~\cite{alur2000discrete, baier2008principles} as well as approximate abstractions~\cite{GIRARD2011568} for a variety of discrete and continuous systems in both deterministic~\cite{pola2008approximately, girard2010approximately} and stochastic settings~\cite{bujorianu2005bisimulation, desharnais2002bisimulation, larsen1991bisimulation, strubbe2005bisimulation}. 
Although several methods have been proposed to compute the simulation metric (see~\cite{ABATE20133} and references therein); their contributions present some limitations; in that; 
1) they validate the open-loop behavior of the model against that of the system, which is unnecessary and may lead to quite conservative bounds on the quality of $\abs$ as an abstraction of $\sys$ for the controller synthesis, as one may be able to control the system well with only a very approximate model of the system, 2) restrictive assumptions on the dynamics of the systems are required, and 3) a computational procedure to determine the approximate simulation metric is given only for certain classes of systems.

More recently, \cite{Abate11approximateabstractions} proposed a randomized approach to compute simulation metric that relies on the formulation of the problem as a semi-infinite chance-constrained optimization and computing its solution using the so-called ``sampling and discarding approach"~\cite{campi2009scenario, campi2011sampling}. 
The proposed approach does not require any \textit{a priori} assumption on the system dynamics, and requires only to be able to run multiple executions of both $\sys$ and $\abs$.
Although, routed with strong theoretical guarantees, the simulation metric is computed through a random sampling-based approach, which often suffers from a poor sample complexity.
Moreover, the validation is performed on the open-loop model; therefore, the bounds obtained on the simulation metric can still be quite conservative. 

To overcome these limitations, we propose \metName{} (Context-Specific validaTiOn framework for data-driven Models), a novel framework to quantify the quality of an abstraction for the controller synthesis purposes.
Here, context refers to a specific (potentially continuous) set of control tasks for which we want to synthesize a controller. 
As opposed to the open-loop validation considered in the above mentioned works, we consider the validation of closed-loop systems, where the controller has been designed using $\abs$ for the particular task. 
This allows us to validate only those behaviors of the model that are relevant for the underlying task rather than across all system behaviors. 
Moreover, we propose to use a \textit{context-specific distance measure}, which considers the context while computing the (closed-loop) distance between $\abs$ and $\sys$ instead of simply comparing how close their trajectories are. 
Hence, the context guides both the controller synthesis procedure~(\textit{context-specific controller}) and the distance measure~(\textit{context-specific distance measure}) in our validation framework. 
This subtlety of \metName{} proves to be very important for validating data-driven models where obtaining a good open-loop abstraction across the entire state-space is extremely challenging; thus, such models will hardly pass a typical simulation metric-based validation test, even when the abstraction is good enough for synthesizing a controller for the actual system. 

\metName{} is based on active sampling and does not require any \textit{a priori} knowledge of the system dynamics.
Since each execution on the real system is expensive, we aim to compute the distance between the system and its abstraction with as few samples as possible.
To that end, we propose to use a Bayesian optimization (BO)-based active sampling method to compute this distance, which takes all past executions into account and suggests the next execution that is most likely to maximize the distance function.
This direct maxima-seeking behavior based on the observed distances ensures that our approach is scalable and highly data-efficient compared to random sampling-based approaches, such as~\cite{Abate11approximateabstractions}.
Overall, the proposed framework provides a \textit{practical} and \textit{scalable} scheme to validate data-driven models for controller synthesis.

\section{Problem Statement} \label{sec:formulation}
Let $\sys$ and $\abs$ denote the true system and its discrete-time model (also referred to as \textit{abstraction} hereon) respectively, with state space $\statespace$ and control space $\controlspace$.
We assume that $\abs$ has the same input and output space as $\sys$ but is often simpler and, in general, learned from data. 
Let $\traj_{\sys}(t; \state_0, \ctrl(\cdot))$ denote the system state at time $t$ starting at state $\state_0$ and applying the control sequence $\ctrl(\cdot)$. 
$\traj_{\abs}$ is similarly defined.

Given $\sys$ and its model $\abs$, our goal is to \textit{validate} $\abs$, i.e., we would like to quantify how useful $\abs$ is for designing controllers for $\sys$ to complete tasks in the taskspace, $\taskspace$. 
More specifically, if we design a controller $\cont(\cdot)$ to complete a task $\task \in \taskspace$ using the model, will it perform equally well on the true system? 
Here, we are interested in controlling the system over a discrete-time horizon of $\horizon$; therefore, each task can be thought of as a control objective defined over the horizon, $\horizon$. 
We also assume that the taskspace $\taskspace$ can be parameterized by $\param \in \paramspace$, i.e., for each $\param \in \paramspace$, we have a different instance of a task $\task_{\param} \in \taskspace$.
For example, if we are interested in regulating the system to a desired state over a horizon of $\horizon$, then $\param$ can represent the desired state, and $\paramspace$ can represent the set of all states to which we might want the system to regulate to.

Typically, the quality of a model is quantified by comparing the open-loop trajectories of $\sys$ and $\abs$.
In particular, a distance measure $d(\traj_{\abs}, \traj_{\sys})$ is defined between the trajectories and its maximum is computed over all possible pairs of trajectories, i.e., 
\begin{equation}
    d^* = \max_{\ctrl(\cdot) \in \controlspace^{\horizon}, \state_0 \in \statespace_0} d(\traj_{\abs}(\cdot; \state_0, \ctrl(\cdot)), \traj_{\sys}(\cdot; \state_0, \ctrl(\cdot))).
    \label{eqn:naive_validation}
\end{equation}
$d^*$ in \eqref{eqn:naive_validation} is also called simulation metric between $\sys$ and $\abs$.
Intuitively, if $d^*$ is small, then the trajectories of the actual system and its abstraction are close to each other, and we can quantify $\abs$ as a ``good" or \textit{validated} model of $\sys$. 
However, if $d^*$ is large, we explicitly have an initial state and a control sequence for which the evolved pair of trajectories will be very different, and we say that $\abs$ doesn't approximate $\sys$ well.

However, the notion of distance in \eqref{eqn:naive_validation} is too general; in that, it measures the distance across all possible control laws without explicitly considering the tasks of interest.
Often, the goal of obtaining an abstraction is to synthesize a controller for the system and hope that it will work well on the actual system. 
The open-loop distance measure in \eqref{eqn:naive_validation} does not capture this intended use of an abstraction.
Moreover, it has been shown in~\cite{Bansal2017a} that a ``good" model for controller synthesis can be very different from a model that mimics the open-loop behavior of the true system well.
Thus, a small open-loop distance measure in \eqref{eqn:naive_validation} is not even necessary for a model to be a good abstraction of the system.

This motivates us to consider a context-specific validation framework, which measures the closed-loop distance between $\sys$ and $\abs$.
The context here is defined by the taskspace, $\taskspace$, which guides two key elements in our framework: \textit{context-specific controller} and \textit{context-specific distance measure}.
In contrast to open-loop control, the context-specific controller refers to a controller designed for the specific task instance using $\abs$, and is represented as $\cont(\state_t;\task_{\param}, \abs)$.
For example, $\cont$ can be designed by associating a cost function with each task (typical in robotics) and solving an optimal control problem.
This mimics the intended use of an abstraction, i.e., to design a controller for the actual system for the tasks at hand.
We succinctly represent $\cont(\state_t;\task_{\param}, \abs)$, as $\cont_{\param}$ to make the dependence on $\param$ explicit.
While it is obvious that $\cont_{\param}$ differs based on the task (and context in general), we claim that the distance measure used to validate an abstraction should also depend on the context at hand.
Often, the entire trajectory of the system is not of interest for the task at hand; in such cases, one should use a distance measure that considers only the relevant part(s) of the trajectories.
For example, for regulation tasks, one may care only about how close the final state of the model is to the desired state, and an appropriate distance measure should only compare the final state of $\abs$ and $\sys$ rather than computing the distance between the full trajectory.
Thus, we propose to validate an abstraction based on the maximum context-specific distance, $\dist^*(\abs_{\cont}, \sys_{\cont})$, between the closed-loop $\sys$ and $\abs$,
\begin{equation}
    \dist^*(\abs_{\cont}, \sys_{\cont}) = \max_{\param \in \paramspace, \state_0 \in \statespace_0} \dist(\traj_{\abs}(\cdot; \state_0, \cont_{\param}), \traj_{\sys}(\cdot; \state_0, \cont_{\param})),
    \label{eqn:max_dm}
\end{equation}
where $\abs_{\cont}$ and $\sys_{\cont}$ denote the closed-loop abstraction and system respectively.
Thus, we aim to validate a system model \textit{given} a model, a control scheme and a control horizon.
For simplicity, we denote $\dist(\traj_{\abs}(\cdot; \state_0, \cont_{\param}), \traj_{\sys}(\cdot; \state_0, \cont_{\param}))$ as $\dist(\param)$ here on to make the dependence on $\param$ explicit.

\begin{definition}[Validation]
We say a model $\abs$ for a system $\sys$ is \textit{validated} for a context $\taskspace$, if the context-specific distance $\dist^*$ is below a pre-defined threshold $\tau$, i.e., $\dist^* \leq \tau$.
\end{definition}

Intuitively, $\tau$ quantifies the maximum acceptable difference between a system and its abstraction. 
Our goal in this paper is to check whether $\abs$ validates $\sys$ or not, for which we need to compute $\dist^*$.
Since $\sys$ is unknown, the shape of the objective function, $\dist(\param)$, in \eqref{eqn:max_dm} is unknown. 
The distance is thus evaluated empirically for each $\param \in \paramspace$, which is often expensive as it involves conducting experiments on the real system. 
Thus, the goal is to solve the optimization problem in \eqref{eqn:max_dm} with as few evaluations as possible. 
In this paper, we do so via a BO-based active sampling method.
\section{Running Example} \label{sec:running_example}
We now introduce a canonical control example, the inverted pendulum, that we will use throughout the paper to illustrate our approach.
It has two states, $x = (\theta, \dot{\theta})$, the angle of rotation, $\theta$; and, angular velocity, $\dot{\theta}$. The dynamics of the system, $\sys$, are given by, 
\begin{equation}
    \label{eqn:inv_pend}
    \frac{m l^2}{3} \ddot{\theta}(t) = \frac{mgl}{2}\textit{sin}(\theta(t)) + u(t)
\end{equation}
where $m$ is the mass of the pendulum, $g$ is the acceleration due to gravity, $l$ is the length of the rod. In~\eqref{eqn:inv_pend}, $\theta$ is the angle of the pendulum measured counter-clockwise from the upward vertical and $u$ is the external torque applied. The state space and control space are, $\statespace = [-\pi, \pi) \times \reals$ and $\controlspace = \reals$ respectively.  
We would like to design a state-feedback controller, $\cont(\theta, \dot{\theta})$ which attempts to stabilize the pendulum at a desired angle, i.e., $\state_{final} = (\theta_{final}, 0)$, starting at the initial configuration $\state_{init} = (\pi, 0)$, i.e., the bottom-most position.

For simulation purposes, we assume that the actual dynamics model in \eqref{eqn:inv_pend} is unknown, and learn a simple linear model $\abs$ from data using least-squares.
The $\taskspace$ is instantiated by the desired angle we would like to regulate to, $\theta_{final}$. Hence, we have a 1D $\paramspace = [-\pi, \pi)$. For each $\param \in \paramspace$, we design a controller $\cont$ for the abstract model $\abs$. We use finite horizon \textit{Iterative Linear Quadratic Regulator (iLQR)} to design this controller, where iLQR minimizes a quadratic cost function penalizing the deviation from the desired state, $\state_{final}$. 

We use the $l_1$ distance between the final states of the closed-loop $\abs_{\cont}$ and $\sys_{\cont}$ as our distance measure, i.e., 
\begin{equation}
\dist(\param) = \|\traj_{\sys}(\horizon; x_{init}, \cont_{\param}) - \traj_{\abs}(\horizon; x_{init}, \cont_{\param})\|_{1}
\label{eqn:IP_dist}
\end{equation}
Note that other distance measures can very well be used for the validation purposes, but the measure in \eqref{eqn:IP_dist} might be a more meaningful distance measure for a regulation task compared to the distance between the trajectories of the two systems, since we, often, only care about the final state of the system in this scenario, rather than the taken trajectory. 
We will discuss the effect of the choice of the distance measure further in Section \ref{subsec:approach}.
\section{Background}
\subsection{Random Sampling}
\label{sec:random_approaches}
If we do not make any assumptions on $\sys$ and $\abs$, then the associated distance function, $\dist(\cdot)$, can be arbitrary. 
Random sampling-based approaches have been explored for solving~\eqref{eqn:max_dm} in such scenarios. 

Consider an equivalent form of~\eqref{eqn:max_dm},
\begin{equation}
    \dist^{*}(\abs_{\cont}, \sys_{\cont}) = \min_{h \in \reals} h 
    \text{, s.t }  \forall_{\param \in \paramspace}\text{  }\dist(\param) \leq h 
\label{eqn:max_dm_rs}
\end{equation}
\eqref{eqn:max_dm_rs} is a semi-infinite convex optimization problem where the objective and constraints are convex in the optimization variable, $h$. For simplicity, we augment $\paramspace$ with the space of initial states, $\statespace_0$, and consider initial states as a parameter in \eqref{eqn:max_dm_rs}.
We now discuss two random sampling-based approaches to solve \eqref{eqn:max_dm_rs}.

\subsubsection{Scenario-based optimization (SC)}
\label{Sec:SB}
Scenario-based optimization has been introduced in~\cite{calafiore2006scenario} and~\cite{campi2009scenario} to solve 
\eqref{eqn:max_dm_rs} by approximating it as a finite convex optimization problem. 
In particular, $N$ parameters $p_1, \dots, p_N \in \paramspace$ are randomly sampled and the optimization problem in \eqref{eqn:max_dm_rs} is solved for these parameters, i.e., we compute
\begin{equation}
    \hat{\dist}^{SC}(\abs_{\cont}, \sys_{\cont}) := \max_{i \in 1, \dots, N} \dist({\param_i}).
\end{equation}

It is shown in~\cite{campi2009scenario} that given any $\epsilon \in (0,1)$ and $\beta \in (0,1)$, the following statement holds with probability no smaller than $(1 - \beta)$
\begin{equation*}
    \text{Pr}(\param \in \paramspace: \dist(\param) > \hat{\dist}^{SC}) < \epsilon ~\text{if}~ N \geq \mathcal{O}(\frac{1}{\epsilon} \log(\frac{1}{\beta})).
\end{equation*}
Intuitively, the above statement provides a probabilistic bound on the volume of the parameter space where the distance is greater than $\hat{\dist}^{SC}$. 
Simply put, as $N$ increases, the volume of $\paramspace$ where the distance is greater than $\hat{\dist}^{SC}$ decreases at the rate of $1/N$.
It is also interesting to see that this rate is independent of the dimension of $\paramspace$.
However, the scenario-based approach does not provide any information on the gap between our current estimate $\hat{\dist}^{SC}$ and the global optimum $\dist^*$, which can be arbitrary in general.
In contrast, our approach provides an explicit bound on this gap, as well as its rate of convergence to zero, as shown in Sec. \ref{subsec:theo_results}.

\subsubsection{Sampling and discarding approach (SD)}
Another approach to solving \eqref{eqn:max_dm_rs} is to reformulate it as a chance-constrained optimization problem, 
\begin{equation}
        \min_{h \in \reals}h
        \text{, s.t }  \text{Pr}( \param \in \paramspace: \dist(\param) > h) \leq \epsilon,
    \label{eqn:max_dm_ccp}
\end{equation}
and solve using sampling-based methods~\cite{campi2011sampling}.
In particular, $N$ parameters $\param_1, \dots, \param_N \in \paramspace$ are sampled randomly and an algorithm $\mathcal{A}$ is used to discard $k$ of these parameters to compute
\begin{equation}
    \hat{\dist}^{SD}(\abs_{\cont}, \sys_{\cont})  := \max_{i \in \{1, \dots, N\} - \mathcal{A}(\param_1, \dots, \param_N)} \dist(\param_i).
\end{equation}

The formulation in \eqref{eqn:max_dm_ccp} has also been adopted in~\cite{Abate11approximateabstractions} to compute distance between a system and it model.
While different algorithm $\mathcal{A}$ can be chosen, for the purposes of computing maximum distance between models, we borrow $\mathcal{A}$ proposed in~\cite{Abate11approximateabstractions}, where the authors discard the parameters corresponding to $k$ maximum distances. 

Similar to the scenario-based optimization, it can be shown that given any $\epsilon \in (0,1)$ and $\beta \in (0,1)$, the following statement holds with probability no smaller than $(1 - \beta)$~\cite{campi2011sampling}
\begin{equation*}
    \text{Pr}(\param \in \paramspace: \dist(\param) > \hat{\dist}^{SD}) < \epsilon ~\text{if}~ \sum_{i=0}^{k-1}\binom{N}{i} \epsilon^{i}
     (1- \epsilon)^{N-i} \leq \beta.
\end{equation*}
%
This probability can be interpreted along similar lines as in scenario-based optimization.
In addition, sampling and discarding approach provides probabilistic bounds on the gap from the optimal $h$ in \eqref{eqn:max_dm_ccp} (for details we refer the interested readers to~\cite{campi2011sampling}).
However, this approach also does not provide any information about the gap from the true maximum $\dist^*$.

\subsection{Active sampling}
\subsubsection{Gaussian Process~(GP)}
\label{Sec:GP}
Since we do not know the dynamics of the actual system, the dependence of $\dist$ on the parameters~$p \in \paramspace$ is unknown \textit{a priori}. We use a GP to approximate $\dist$ in the domain $\paramspace$. The following introduction of GPs is based on~\cite{Rasmussen}.

GPs are non-parametric regression method from machine learning, where the goal is to find an approximation of the nonlinear function $\dist : \paramspace \rightarrow \reals$ from a parameter $p \in \paramspace$ to the function value $\dist$. This is done by considering the function values $\dist(p)$ to be random variables, such that any finite number of them have a joint Gaussian distribution. 
For GPs, we define a prior mean function and a covariance function (or kernel), $k(p, p')$ which defines the covariance between the function values $\dist(\param)$ and $\dist(\param^{'})$. 
The prior mean in our case is assumed to be zero without loss of generality. 
In general, the choice of kernel function is problem-dependent and encodes assumptions about the unknown function such as smoothness.
In our experimental section, we use the $3/2$ Mat\`ern kernel where the hyper-parameters are optimized by maximizing the marginal likelihood~\cite{Rasmussen}.

The GP framework can be used to predict the distribution of the performance function $\dist(p)$ at an arbitrary input $\param \in \paramspace$ based on the past observations, $\dataset=\{\param_i,\dist(\param_i)\}_{i=1}^n$.
Conditioned on $\dataset$, the mean and variance of the prediction are
\begin{equation} \label{eq:one-step prediction mean and covariance}
\mu(\param) = {\vec k}\mat K^{-1} {\vec \cost};~~ \sigma^2(\param) = k(\param,\param)-{\vec k}\mat K^{-1}{\vec k}^T,
\end{equation}
where $\mat K$ is the kernel matrix with $K_{ij}= k(\param_i,\param_j)$, $\vec k =[k(\param_1,\param),\ldots,k(\param_n,\param)]$ and $\vec \cost =[\dist(\param_1),\ldots,\dist(\param_n)]$. 
Thus, the GP provides both the expected value of the performance function at any arbitrary point~$\param$ as well as a notion of the uncertainty of this estimate. 


\subsubsection{Bayesian Optimization (BO)}
\label{sec:bayesian_optimization}
In this work we use BO in order to find the maximum of the unknown function~$\dist(\cdot)$. 
BO is particularly suitable for the scenarios where evaluating the unknown function is expensive, which fits our problem in Sec.~\ref{sec:formulation}.
At each iteration, BO uses the past observations $\dataset$ to model the objective function, and uses this model to determine informative sample locations. 
A common model used in BO for the underlying objective, and the one that we consider, are Gaussian processes (see Sec. \ref{Sec:GP}). 
Using the mean and variance predictions of the GP from \eqref{eq:one-step prediction mean and covariance}, BO computes the next sample location by optimizing the so-called acquisition function, $\acqfunc{\cdot}$.
Different acquisition functions are used in literature to trade off between exploration and exploitation during the optimization process.
The acquisition function that we use here is the GP-Upper Confidence Bound (GP-UCB)~\cite{srinivas2009gaussian}, where the next evaluation is given by $\param^{*} = \argmax_{\param \in \paramspace} \acqfunc{\param}$, 
\begin{equation}
\label{eqn:BO_f_acqu}
\acqfunc{\param} = \mu(\param) + \beta^{1/2} \sigma(\param),
\end{equation}
where $\beta$ determines the confidence interval and is typically chosen as~${\beta_n=2}$.
Intuitively, at each iteration \eqref{eqn:BO_f_acqu} selects the next parameter for which the upper confidence bound of the GP is maximal. 
Repeatedly evaluating the true function~$\dist$ at samples given by \eqref{eqn:BO_f_acqu} improves the GP model and decreases uncertainty at candidate locations for the maximum, such that the global maximum is found eventually.


\section{Active Sampling for Distance Computation}
We now present \metName{}, an active sampling-based framework to compute the context-specific distance between a system and its model.  

\subsection{Approach} \label{subsec:approach}
Our approach is summarized in Algorithm \ref{alg:solution}.
We start with a given model $\abs$ that we want to validate for all possible tasks in the taskspace $\taskspace$, parameterized by $\param \in \paramspace$.
In \metName{}, we model the unknown function $\dist$ as a GP.
Given an initial $\dataset$~(initialized randomly if empty), we initialize our GP~(Line 3). Next we compute the $\param^{*}$ that maximizes the corresponding acquisition function $\acqfunc\param$ (Line 5).
This parameter, $\param := \param^*$ is used to design a controller $\cont(\state_{\time};\task_{\param}, \abs)$ for the model $\abs$ (Line 6).
This controller should be the same as the one that the system designer would synthesize for the actual system based on $\abs$.
Typically, this is done by solving an optimal control/reinforcement learning problem that minimizes some task-specific cost function $\totalcost$ subject to $\abs$. 
We then apply the controller in a closed-loop fashion to both the abstract model, $\abs$, and the actual system, $\sys$; and the corresponding state-input trajectories are recorded (Lines 7 and 8). 
These trajectories are used to compute the distance $\dist(\param)$ (Line 9).
The GP is updated based on the collected data sample $\{\param, \dist(\param)\}$ (Line 10) and the entire process is repeated until the $\param^*$ corresponding to the maximum distance is found.

%
\begin{algorithm}[t]
	\DontPrintSemicolon
	\caption{\metName{} algorithm}
	\label{alg:solution}
	$\dataset$ \hspace{2.9 mm} $\longleftarrow$ if available: $\{\param, \dist(\param)\}$; otherwise, initialized randomly\;
	Prior $\longleftarrow$ if available: Prior of the GP hyperparameters; otherwise, uniform\;	
	Initialize GP with $\dataset$\;	
	\While{\text{optimize}}{	
		Find $\param^* = \argmax_{\param} \acqfunc{\param}$; $\param \longleftarrow \param^*$ \;
		Design $\cont(\state_{\time};\task_{\param}, \abs)$ for the task $\task_{\param}$ \;
		Apply $\cont$ on $\abs$ and record $\traj_{\abs}(\cdot; \state_0, \cont(\cdot))$ \; 
		Apply $\cont$ on $\sys$ and record $\traj_{\sys}(\cdot; \state_0, \cont(\cdot))$ \;
		Evaluate $\dist(\param)$ based on the defined distance measure\;
	    Update GP and $\dataset$ with $\{\param,\dist(\param)\}$
    }
\end{algorithm}
%

Intuitively, \metName{} directly learns the shape of the distance function $\dist(\param)$ as a function of $\param$. However, instead of learning the global shape of this function through random queries, it analyzes the performance of all the past evaluations and by optimizing the acquisition function, generates the next query that is the most likely candidate for the maxima of the distance function. 
This direct \textit{maxima-seeking} behavior based on the \textit{observed distance} ensures that \metName{} is highly data-efficient. 
Equivalently, in the space of all tasks of interest, we efficiently and directly search for the task for which the abstract model performs ``most differently" compared to the actual system. 
For our running example, thus, we are directly seeking the angle which is \textit{hardest} to balance the pendulum at, using a controller designed based on the learned model.

Note that the mean of the GP in our algorithm can be initialized with a prior distance function if some information is known about it. 
This generally leads to a faster convergence. 
When no information is known about the distance function \textit{a priori}, the initial tasks are queried randomly and the corresponding distances are used to initialize the GP. 
Finally, note that \metName{} can also be used when the distance measure is stochastic, which for example is the case when the actual system or the model is stochastic. 
In this case, \metName{} will maximize the expected distance.

There are two natural questions to ask at this point:
\begin{enumerate}
    \item Is the context-specific validation framework in \eqref{eqn:max_dm} any better than the open-loop validation framework in \eqref{eqn:naive_validation}?
    \item Is active sampling approach more sample efficient than scenario-based or sampling and discarding approaches?
\end{enumerate}

\begin{example}
We now provide some insights into above questions using our running example in Sec. \ref{sec:running_example}.
To demonstrate the utility of a context-specific validation framework, we consider the following four scenarios 
\begin{enumerate}
    \item Open-loop control, distance between trajectories~(OLDT): The distance is defined by~\eqref{eqn:naive_validation}, and we measure the open-loop distance between the system and the model. We solve for, 
    \begin{equation*}
        d^* = \max_{u \in \controlspace^{\horizon}, \param \in \paramspace} \|\traj_{\abs}(\cdot;x_{init}, u) - \traj_{\sys}(\cdot;x_{init}, u)\|_{\infty}
    \end{equation*}
    \item Context-specific controller, distance between trajectories~(CCDT): Here, the context is used only to design the controller $\cont$ while the distance is defined between the trajectories using infinity norm. We solve for,
    \begin{equation*}
        d^* = \max_{\param \in \paramspace} \|\traj_{\abs}(\cdot; x_{init}, \pi_{\param}) - \traj_{\sys}(\cdot; x_{init}, \pi_{\param})\|_{\infty}
    \end{equation*}
    \item Open-loop control, Context-specific distance metric~(OLCD): Here the context is considered only to define $\dist$, i.e, the distance measure in~\eqref{eqn:IP_dist}, but an open-loop controller is used. The distance between the two models is given by, 
    \begin{equation*}
        d^* = \max_{u \in \controlspace^{\horizon}, \param \in \paramspace} \|\traj_{\abs}(H; x_{init}, u)- \traj_{\sys}(H; x_{init}, u)\|_{1}
    \end{equation*}
    \item Context-specific validation~(\metName{}): Here we use our proposed approach, i.e., the context is considered while designing the controller, $\cont$, as well as in the distance measure, $\dist$, and solve for 
        \begin{equation*}
        d^* = \max_{\param \in \paramspace} \|\traj_{\abs}(H; x_{init}, \pi_{\param}) - \traj_{\sys}(H; x_{init}, \pi_{\param})\|_{1}
    \end{equation*}
\end{enumerate}

In~\cref{fig:compare_frameworks} we plot the distance function for each scenario as a function of the task parameter, $\param := \theta_{final} \in [-\pi, \pi]$. 
The distance profiles for OLDT~(shown in green) and OLCD~(shown in red) are independent of $\theta_{final}$, as the control law is open-loop and does not consider the task explicitly.
Moreover, the distance for any given $\theta_{final}$ is higher than that of CCDT~(shown in orange) or \metName{}~(shown in blue), because we are comparing the abstraction and the system for all possible finite horizon control laws. 
Although OLDT/OLCD give the maximum distance between the trajectories of $\abs$ and $\sys$, they convey no information about how useful $\abs$ is for designing controllers to regulate the inverted pendulum to the desired angle.
This illustrates the utility of comparing the closed-loop models $\abs_{\cont}$ and $\sys_{\cont}$ rather than the open-loop models. 
This is an \textit{important} observation: since $\abs$ is learned for synthesizing $\cont$, we need to analyze how well $\cont$ behaves on the real system $\sys$, and not necessarily all possible control laws.  
Indeed, we see that CCDT/\metName{} has a much lower distance profile compared to OLDT and OLCD, implying that the learned model may not necessarily have the same open-loop trajectories as that of the actual system, but their closed-loop trajectories are relatively similar. 
%
\begin{figure}
    \centering
    \includegraphics[width=\columnwidth]{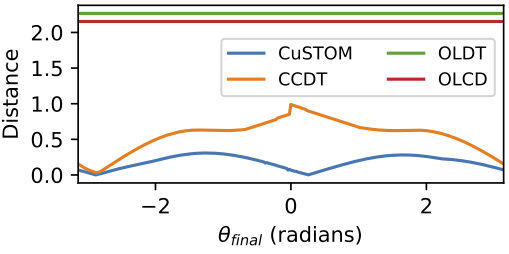}
    \caption{Inverted Pendulum. The $x$-axis and $y$-axis show the parameter space, $\theta_{final} \in [-\pi, \pi]$, and the distance between $\abs$ and $\sys$ respectively. 
    The distance profiles of the open-loop frameworks (shown in red and green) are independent of $\theta_{final}$ and compute the distance across all finite horizon control laws.
    Using a context-specific controller (shown in blue and orange), reduces the maximum distance significantly, indicating that the learned model may not mimic the actual system everywhere, but their closed-loop behavior is very similar for the tasks of interest.
    }
    \label{fig:compare_frameworks}
\end{figure}

Furthermore, a good distance measure should accurately capture the intent of the system designer.
For example, expecting the trajectories of the closed-loop systems to be close maybe too strict for regulation purposes.
Instead, one might be interested in how close the final states of $\abs_{\cont}$ and $\sys_{\cont}$ are.
A small distance will then imply that we can expect the actual system to get close to the desired angle, given that the model can be regulated to the desired angle.
Indeed, comparing \metName{} and CCDT indicates that the trajectories of $\abs_{\cont}$ and $\sys_{\cont}$ do not align exactly, but reach roughly the same final state. 
Fig. \ref{fig:compare_trajs_IP} illustrates the difference in the trajectories of $\abs_{\cont}$ and $\sys_{\cont}$ for $\theta_{final} = 0.04$, corresponding to the maximum of CCDT.
The controller designed on the abstract model leads to a larger undershoot on the actual system but leads to the same final angle, resulting in a larger distance as per CCDT, but a smaller distance as per \metName{}.
In general, the choice of the distance measure is subjective, but the validation threshold should be chosen keeping in mind the chosen distance measure.
For example, it will be too strict to pick the same validation threshold for CCDT and \metName{} in this case.
\begin{figure}
    \centering
    \includegraphics[width=\columnwidth]{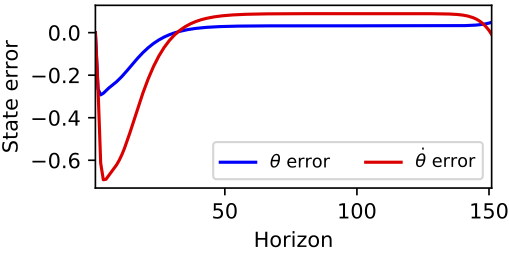}
    \caption{Inverted Pendulum. The error between the closed-loop trajectories of $\abs_{\cont}$ and $\sys_{\cont}$ as horizon progresses. 
    The controller is designed using $\abs$ for $\theta_{final} = 0.04$.
    Note that even though the controller leads to different undershooting behavior on $\abs_{\cont}$ and $\sys_{\cont}$, they finally reach the same configuration. 
    Thus, the choice of the distance measure affects the distance between $\abs_{\cont}$ and $\sys_{\cont}$, and an appropriate measure should be chosen that captures the intent of the designer.
    }
    \label{fig:compare_trajs_IP}
\end{figure}
%
\end{example}

\begin{example}
We next compare the effectiveness of the three methods, scenario-based optimization~(SC), sampling and discarding approach~(SD) and active sampling using BO (\metName{}), to solve~\eqref{eqn:max_dm}.
\cref{fig:compare_techniques} compares the error in the maximum distance estimation across the three methods as a function of the number of parameter samples. 
The first five samples used to initialize all the three methods are the same. 
It takes \metName{} (the Blue curve) only \textit{three} samples to reach within the $1\%$ of the true distance, as opposed to SC (the Orange curve) which is able to reach within $2.5\%$ of the true distance only after $60$ samples, which supports our claim that modeling $\dist$ as a GP and using BO converges to the true distance faster. 
Further, we notice that SD (the Green curve) does not reach close to the true distance even after 100 samples. 
This is not surprising as SD considers a chance constrained formulation and discards $k> 0$~(in our case $k=5$) maximum samples.
\begin{figure}[t]
    \centering
    \includegraphics[width=\columnwidth]{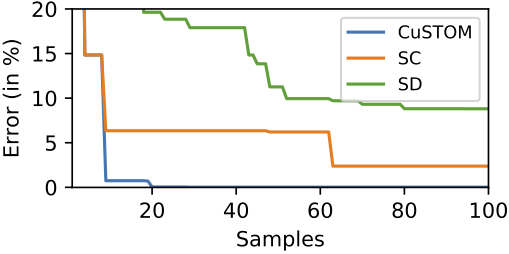}
    \caption{Inverted Pendulum. The $x$-axis and $y$-axis show the number of samples and the percentage error in the distance estimation respectively. While the error for BO (shown in Blue) converges to 0 very quickly~(in 20 samples), scenario-based optimization~(SC) (shown in Orange) is not able to find the true maximum even after 100 samples. Both techniques outperform sampling and discarding~(SD) approach (shown in Green), which does not seem close to the true distance even after 100 samples.}
    \label{fig:compare_techniques}
\end{figure}

The converged GP for \metName{} and the uncertainty associated with it is shown in Fig.~\ref{fig:converged_GP}.
The true distance function is shown in red (computed by gridding the parameter space). 
While the GP does not capture the entire distance function within the $95\%$ confidence interval~(the red line is not completely contained in the grey uncertainty region), it captures the region around maximum very well, which is sufficient for BO.  
In fact, this characteristic of BO precisely makes it more sample efficient compared to some of the other approaches. 

Additionally, the system designer can use the GP itself in several different ways to get important insights about the learned model.
First, it tells the designer that the maximum distance which is $0.3$ radians occurs when we want to regulate the system to $\theta_{final} = 2$ radians. 
In particular, we can expect the actual system angle to be in the range $[1.7, 2.3]$ radians, if we use the controller designed on $\abs$\footnote{This assumes that we can regulate the model to the desired angle.}.  
This distance should be compared with the predefined tolerance threshold to validate the model, which is also the main focus of the paper. 
In particular, if a distance of $0.3$ radians is within the tolerance the designer is willing to accept, we can conclude that the learned linear model is an acceptable model for the inverted pendulum for controller synthesis. 
However, if it is beyond the acceptable tolerance, we can precisely tell the designer for what regulation tasks, the model fails. 
Note that this is different from saying that the linear system is sufficient for regulating the inverted pendulum, since we do not analyze how good the controller is at \textit{completing} the task. 
It merely suggests that the controllers designed for regulation purposes behave similarly with the model, $\abs$, and actual system, $\sys$. 
In that aspect our validation framework is a tool in the overall control analysis procedure (see Remark \ref{remark:val_aspects} for further discussion).
\begin{figure}[t]
\centering
    \includegraphics[width=\columnwidth]{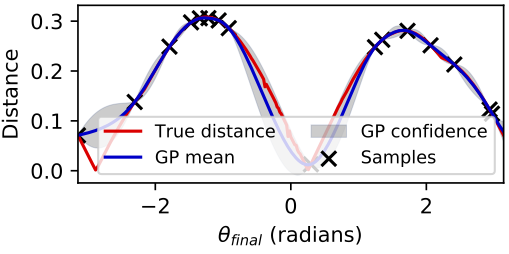}
    \caption{Inverted Pendulum. The $x$-axis and $y$-axis are the parameter space and the distance function $\dist(\param)$ respectively. The true distance function is shown in red, the mean of the GP is shown in blue and the shaded grey region is the associated uncertainty.
    It suffices for the GP to capture the true distance function around the maximum of the true distance function, and not necessarily over the entire parameter space.}
    \label{fig:converged_GP}
\end{figure}

Second, it indicates that the distance profile is asymmetric. 
This reflects a potential bias in the training data while learning $\abs$ since the inverted pendulum is symmetric around the bottom-most position. 
Furthermore, $\abs_{\cont}$ performs relatively poorly around $\pi/2$ and $-\pi/2$, which suggests that $\abs$ needs to be refined further by adding more samples from this region in the training data.
Thus, using the GP, we can obtain some tangible information about the regions in which the model can be refined further to improve the closed-loop performance on the actual system, something that random sampling-based approaches are unable to provide.
\begin{remark} \label{remark:val_aspects}
In general, there are two, perhaps equally, important aspects to a context-specific validation:
\begin{itemize}
    \item Aspect-1: How well a \textit{model} can perform the tasks in $\taskspace$?
    \item Aspect-2: How well a model can mimic the \textit{actual} system behavior on these tasks?
\end{itemize}
Aspect-1 checks if model itself can perform the tasks to a satisfactory level, i.e., verify~(or test) the model can perform the task.
Aspect-2 checks if we can expect the same performance to transfer to the actual system.
Thus, to make sure that the designed controller performs well on the desired tasks, it is important to test and validate a model along both aspects. 
In this paper, we are concerned with efficiently validating the model along Aspect-2; nevertheless, a model should be first tested along Aspect-1, and only then along Aspect-2.
In fact, if we are not able to verify the model along Aspect-1, validating it along Aspect-2 provides very limited information.
\noindent Several methods, such as simulation-based falsification~\cite{donze2010breach, s_taliro}, have been proposed in literature to test a model against a high level specification along Aspect-1. 
These methods can easily be used along with \metName{} to complete the validation process.

It is also important to note that verification or testing of models is purely simulation-based and does not require any real execution on the system. 
Consequently, it is often much ``cheaper" compared to validating a model along Aspect-2.
Finally, these two aspects can also be combined in a single distance measure. For instance, for regulation tasks, we can validate an abstraction using the following distance measure
\begin{equation*}
    d^* = \max_{\param \in \paramspace} \|\traj_{\sys}(H; x_{init}, \pi_{\param}) - \param\|_{1}.
\end{equation*}
The above distance measure directly validates a model based on the performance of the synthesized controller on the actual system for the task it was designed for.
\metName{} can then be similarly used to validate a model using this distance measure.
However, validating a model in two separate stages helps in identifying the concrete reasons for failing the validation, which can be instrumental in improving~(or refining) the model further. 
\end{remark}
\end{example}

\subsection{Theoretical Analysis} \label{subsec:theo_results}


We now discuss the rate of convergence of \metName{}. 
Global optimization is a difficult problem without any assumptions on the objective function  $\dist$. 
The main complicating factor is the uncertainty over the extent of the variations of $\dist$.
For example, $\dist$ could be a characteristic function, which is equal to 1 at $\param_m$ and 0 elsewhere, and none of the methods we mention here can optimize this function without exhaustively searching through every point in $\paramspace$.
The way a large number of global optimization methods address this problem is by imposing some prior assumption on how fast the objective function $\dist$ can vary. 
A common assumption made in the optimization literature is Lipschitz continuity of $\dist$ in $\param$ \cite{de2012exponential}, which essentially limits the rate of change of $\dist$. 
However, a single Lipschitz constant for the entire function might be too conservative \cite{bubeck2011x} for the exploration-exploitation trade-off required for a data-efficient global optimization. 
One way to relax these hard Lipschitz constraints is by imposing a GP prior on the function, as we do in \metName{}. Instead of restricting the function from oscillating too fast, a GP prior requires those fast oscillations to have low probability. 
Under a GP prior, it can be shown that \metName{} converges to $\dist^*$ at a sub-linear rate.
%
%
\begin{proposition} \label{prop:prob_gua}
Assume $\dist \sim GP(0, k(\param, \param^{'}))$ with a Mat\'ern 3/2 kernel. Select $\epsilon \in (0, 1)$ and sample $p_n \in \paramspace$ by maximizing~\eqref{eqn:BO_f_acqu}. Then,
\begin{equation}
     \text{Pr}(\dist^* > \hat{\dist}^{BO}(n) + r_n\; \forall n \geq 1) < \epsilon 
\end{equation}
where $r_n = C\bigg(\sqrt{\frac{|\paramspace|}{n^{\frac{3}{3+|\paramspace|(|\paramspace|+1)}}}}\bigg)$ for some constant $C$, $|\paramspace|$ is the dimension of $\paramspace$, $n$ is the number of iterations of BO, $\dist^*$ is the true maximum distance, and $\hat{\dist}^{BO}(n)$ is the maximum returned by BO after $n$ iterations. 
\end{proposition}
Intuitively, Proposition~\ref{prop:prob_gua} quantifies how far our current best guess, $\hat{\dist}^{BO}$, is from the true maximum $\dist^*$.
Furthermore, it claims that this gap between our current best guess and the true maximum vanishes quickly as the number of samples increases. 

Note that the guarantees provided by Proposition~\ref{prop:prob_gua} are much stronger than the that by the scenario-based or sampling and discarding approaches; not only does it quantify how far the true maximum is from our current estimate but also provides a rate of convergence.

\begin{corollary} \label{cor:prob_val}
If there exists $n > 0$ such that $\hat{\dist}^{BO}(n) < \tau - r_n$, then with probability atleast $1 - \epsilon$ we have $\dist^* < \tau$ and the model $\abs$ for a system $\sys$ has been \textit{validated}. 
\end{corollary}
Corollary~\ref{cor:prob_val} is a direct consequence of Proposition~\ref{prop:prob_gua} and provides a probabilistic validation certificate for $\abs$.
The proof for Proposition~\ref{prop:prob_gua} and Corollary~\ref{cor:prob_val} can be found in the appendix.

\begin{remark}
Even though Proposition~\ref{prop:prob_gua} is specific to Mat\'ern32 kernel, the convergence rates for different kernels can be similarly obtained (see the proof of Proposition~\ref{prop:prob_gua} for more details). 
\end{remark}

A natural question to ask is when can we use a GP prior on the underlying distance function? 
Even though there exist formal conditions in literature for when a GP prior can be used for a function, they are often hard to verify beforehand, especially when the function is unknown \textit{a priori}~\cite{Rasmussen}. 
Nevertheless, GP priors have been successfully used in machine learning and control theory to efficiently model and control unknown systems, primarily because the flexibility they provide for capturing different underlying functions by an appropriate choice of kernel.
In particular, depending on what is known about the distance measure $\dist$ beforehand, an appropriate kernel can be used to increase the likelihood of capturing the true distance function with a GP. 
For example, if the true system, $\sys$, and the abstract system, $\abs$, satisfy the standard results of existence and uniqueness of state trajectory \cite{coddington1955theory}; and the controller, $\cont$, designed to achieve a task, $\task_{\param} \in \taskspace$, is Lipschitz bounded in the parameter $\param$; then the underlying closed-loop systems, $\sys_{\cont}$ and $\abs_{\cont}$ are continuous in parameters $\param$. 
Any continuous distance measure, $\dist$ defined on $\sys_{\cont}$ and $\abs_{\cont}$ is then continuous and deterministic in $\param$.
In such a case, a Mat\'ern Kernel (with $\nu = 1.5$) can be used, which only assumes the continuity of the underlying function \cite{Rasmussen}. 
Similarly, if $\dist$ is also differentiable, alternative kernels can be used to capture it efficiently under a GP prior.
\section{Case Studies} \label{sec:case_studies}
We now apply \metName{} on a series of systems and learned models. 
As baselines, we compare \metName{} to random sampling-based approaches: the scenario-based and the sampling and discarding-based approaches.
\subsection{Dubins Car} \label{subsec:dubins}
For our first simulation, we vary the taskspace, $\taskspace$, i.e., ``context", and show how our technique can be used across different contexts. 
We consider the task of controlling, a three dimensional non-linear Dubins car to follow a desired trajectory. 
The dynamics of the system are given by, 
\begin{equation} \label{eqn:dubins_dyn}
    \dot{y} = v \cos{\phi}, \text{  } \dot{z} = v \sin{\phi}, \text{  }\dot{\phi} = \omega
\end{equation}
where $\state :=(y, z, \phi)$ is the state of the system; with position, $(y, z)$, and heading, $\phi$; and control inputs are the velocity $v$ and turn rate $\omega$.
We would like $\sys$ to follow the desired trajectory, $\hat{\traj}$, over the horizon, $\horizon =100$, starting from the initial state, $\state_{init} = (0, 0, 0)$,

\begin{equation*}
    \hat{\traj}(t; \state_{init}, \param) =\left(\begin{matrix}
    \frac{t\cdot a_0}{\horizon}\\
    \sum_{j=1}^{10} a_j \sin(\frac{2 \pi  j \cdot y(t)}{a_0}) \\
    0\end{matrix}\right), t \in [0, \cdots, \horizon]
\end{equation*}
Thus, $\hat{\traj}(\cdot)$ is a sum of ten sine waves in the $yz$ plane. 
The task parameters are, $a_0\in [0, 2]$, the desired final $y$ position ; and $\{a_1, \cdots, a_{10}\} \in [0,1]^{10}$, the set of amplitudes corresponding to the different sine components. 
The combined parameter space is 11D, $\paramspace = [0, 2] \times [0, 1]^{10}$, and every instance $\param \in \paramspace$ generates a different trajectory, $\hat{\traj}(\param)$.   
    
We assume that the dynamics in \eqref{eqn:dubins_dyn} are unknown, and learn a linear dynamics model, $\abs$, from random data gathered by simulating the system. 
For each trajectory tracking task, $\task_{\param} \in \taskspace$, we use a quadratic cost penalizing the distance between the current state, $x(t)$, and the desired state, $\hat{\traj}(t)$, as the cost function to design a LQR controller, $\cont_{\param}$. The context-specific distance measure, $\dist$, is the $l_2$ distance between the trajectories produced by $\sys_{\cont}$ and $\abs_{\cont}$, i.e.,
\begin{equation*}
    \dist =  \frac{1}{\horizon}\|\traj_{\sys}(\cdot; \state_{init}, \cont_{\param}) - \traj_{\abs}(\cdot; \state_{init}, \cont_{\param})\|_2 ,
\end{equation*}
but other distance measures can very well be used.
We ran the simulation for 10 different trials. In each trial, we generate new training data and learn a linear model from it. We compare the maximum distance found by \metName{}, scenario optimization~(SC) and sampling and discarding~(SD). 
We implement \metName{} as a python package~(will be published after acceptance). 
For BO, we use the python package GPyOpt~\cite{gpyopt2016}.
We do not compute the true distance, as exploring the parameter space, $\paramspace$, becomes a near impossible task for large dimensions. 

\begin{figure}
    \centering
    \includegraphics[width=\columnwidth]{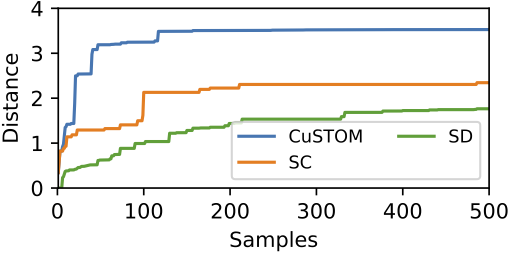}
    \caption{Dubins Car: The $x$-axis and $y$-axis represents the number of samples and maximum distance among the samples, respectively.
    \metName{} converges to its maximum distance much faster than random sampling-based approaches. 
    Moreover, it converges to a higher distance, indicating that SC and SD underestimate the true distance in this case.}
    \label{fig:dubins_car}
\end{figure}

We plot the medians of the distance across the trials for each technique in~\cref{fig:dubins_car}. We see that for a given number of samples, the maximum distance found by \metName{} is higher than that of SC and SD, indicating that the random sampling-based approaches underestimate the true distance between the system and the model in this case.
In fact, if the validation threshold is set somewhere between $2$ to $3$ metres, SC and SD will provide a \textit{false} validation certificate for the learned model, which can be problematic for the safety-critical systems.  

Furthermore, \metName{} converges to its maximum distance in $100$ samples, while SC and SD are not able to find this even after $500$ samples. 
The distance found by \metName{} has a number of implications to the control designer. 
It tells that the designer can expect a maximum average deviation of $3.5$ metres from the trajectory obtained by utilizing $\abs$ for designing $\cont$, and if it is greater than the tolerance of error he is willing to accept then he should refine the model. 
Moreover, our framework provides the designer the trajectory which leads to this error, which he can use to refine the model further.


\subsection{2-DoF Robotic Arm} \label{subsec:reacher}
This simulation is inspired from the manipulation tasks in robotics. 
We consider a two degree-of-freedom (DoF) robotic arm, which is operating on a 2D table.
The goal is to regulate the end-effector of the arm to a desired target position on the table starting from the initial extended arm configuration, i.e., zero joint angles. 

For simulating the actual system, we use the MuJoCo \cite{todorov2012mujoco} physics engine. 
The state of the system, $\state$, is given by the two joint angles, $(\theta_1, \theta_2)$, the corresponding angular velocities, $(\dot{\theta_1}, \dot{\theta_2})$, and the position of the end-effector, $(y, z)$.
The control inputs are the torques for the two motors at the joints.
We assume that we know the state and control of the system, but we do not have access to a dynamics model of the system.
Instead, we learn a dynamics model using the data from the system and use \metName{} to quantify the quality of the learned model for the desired regulation tasks.
To learn the dynamics model, we use a feedforward neural network (FNN) with 2 hidden layers, 8 nodes per layer and ReLU activation function. 
The NN was trained in a supervised fashion with the state-control pair as its input and the next state as the output.

The $\taskspace$ is parameterized by the desired target position of the end-effector, $(y_{final}, z_{final})$, i.e., $\paramspace$ is different positions on the 2D table.
The $l_2$ distance between the final position of the actual arm and its model is used as the distance measure, $\dist$.
A quadratic cost penalizing the distance between the end-effector position, $(y(t), z(t))$, and the desired position, $(y_{final}, z_{final})$, is used as the cost function to design the LQR controller. 

\begin{figure}
    \centering
    \includegraphics[width=\columnwidth]{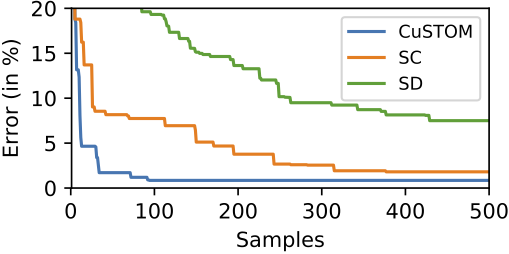}
    \caption{Robotic Arm: the sample complexity for different approaches to reach within a desired percentage of the true distance. \metName{} is significantly more sample-efficient compared to the random sampling-based approaches.}
    \label{fig:reacher_sample_complexity}
\end{figure}

We ran the simulation for 10 different trials, where the training data, and hence the learned model, was different for each trial.
For each trial we also computed the true distance between the system and the learned model by an exhaustive search over $\paramspace$, i.e., across all possible desired positions on the table. 
The median sample complexity across these trials for different approaches is plotted in \cref{fig:reacher_sample_complexity}.
As evident from the figure, \metName{} significantly outperforms all other baselines and reaches within $2\%$ of the true distance in less than 50 executions. 

\begin{figure}
    \centering
    \includegraphics[width=\columnwidth]{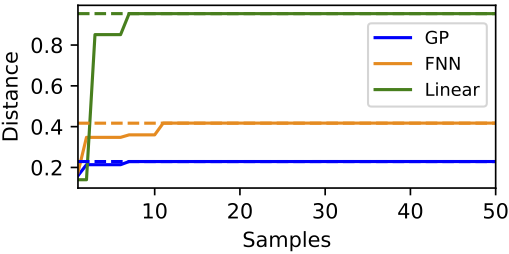}
    \caption{Robotic Arm: Obtained distances for the different abstractions (a Gaussian process (GP), a feed-forward neural network (FNN) and a linear model (LTI)) of a 2-DoF robotic arm. 
    \metName{} can also be used to compare different abstractions of a system, based on the obtained closed-loop distance bound.
    In this case, a GP best mimics the closed-loop behavior of the arm.
    }
    \label{fig:reacher_model_comparison}
\end{figure}

Our framework can also be employed to compare different abstractions of a system.
To illustrate this, we train three different dynamics models for each of the 10 trials above: a FNN model as before, a GP model and a linear model.
\metName{} was then used to validate each of these models.
The results for a particular trial are shown in Fig. \ref{fig:reacher_model_comparison}.
As evident from the figure, \metName{} is able to find the true distance (dashed lines) within a few samples across all models. 
Moreover, based on the observed distances it is clear that the GP model performs best, in the sense that the controller designed on the GP model takes the actual system closest to the desired position, when deployed on the actual system.
In this case, the linear model is the worst abstraction of the real system.
In general, the quality of a data-driven abstraction depends on the training data, training process, function approximator, etc.  
However, such considerations are beyond the scope of this work.
Here, we aim to provide a certificate for an abstraction \textit{given} the abstraction.   
It is also important to note that \metName{} can similarly be used to compare different controllers.

\subsection{Linear Systems} \label{subsec:lin_sys}
The main objective of this simulation is to demonstrate the effect of dimensionality of the task space, $\taskspace$, on the sample complexity of \metName{} and the baselines, SC and SD.

For this purpose, we consider linear systems, $\dot{x} = Ax + Bu$, with state-space dimension ranging from 1D to 10D.
The parameters of the system matrices $(A, B)$ have been chosen in random for each system.  
Corresponding to each system, we consider a linear abstraction whose $(A, B)$ matrices have also been chosen in random. 
For each linear system, we are interested in regulating the system state from a desired initial state, $x_{init}$ to a desired goal state, $x_{final}$. Hence, our $\taskspace$ is parameterized by $\paramspace$ which contains both the initial state and the goal state, $\param = (x_{init},x_{final}) \in \paramspace$.
Thus, our $\paramspace$ is 2D for a 1D linear system, 4D for a 2D linear system, and so on.

A linear feedback controller is used to achieve the desired regulation, whose parameters are also chosen in random.
In particular, the control applied at state $x$ is given by $u(x) = K(x - x_{final})$, where $K$ is the feedback matrix and $x_{final}$ is the desired final state.
In practice, the abstraction as well as the controller are designed carefully based on the data and the task at hand, but in this simulation, our focus is not on the model design but on the relative sample complexity. 

\begin{figure}
    \centering
    \includegraphics[width=\columnwidth]{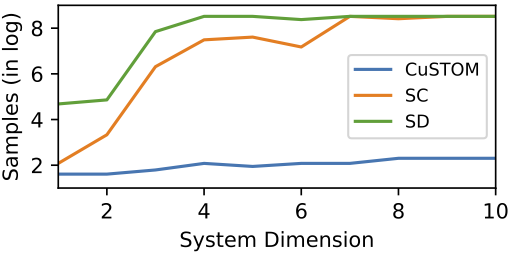}
    \caption{Effect of dimensionality: The sample complexity for different approaches to reach within $5\%$ of the true distance vs. the dimension of the task space. The high sample complexity of random sampling-based approaches renders them impractical for validating abstractions of the real-world autonomous systems.}
    \label{fig:linear_sample_complexity}
\end{figure}

The $l_1$ distance between the final state of the system and the abstraction is used as the distance measure for validation.
For comparison purposes, we also compute the true distance between the model and the abstraction.
This can be done in the closed form for the linear systems, as the distance function, $\dist(\param)$, is also linear in the task parameters, $\param \in \paramspace$.
The median results for the number of samples necessary to reach within the $5\%$ of the true distance are shown in Figure \ref{fig:linear_sample_complexity}.
Even though the number of required samples increases with the dimension of the task space for all the three methods, the increment is very modest for \metName{}. 
This is not surprising because of the sub-linear dependence of sample complexity on the task space dimension for \metName{} (see Proposition \ref{prop:prob_gua}).
However, the sample complexity increases drastically for the two baselines with the dimension of the task space, indicating their impracticality to be used for validating abstractions of real-world systems. 
\section{Conclusion} \label{sec:conclusion}
Validating data-driven models before they are deployed to synthesize a controller for the actual system is a challenging but an important problem. 
In this paper, we present a context-specific validation framework for data-driven models that only validates those behaviors of the model that are relevant for the tasks at hand. 
The proposed framework is based on active-sampling and does not require any \textit{a priori} knowledge of the system dynamics.
The context-specific nature of the framework along with the active sampling ensures that the validation can be performed in a sample efficient manner, and have the potential to be of practical use for real-world systems.

There are several interesting future directions that emerge out of this work. 
First, it would be interesting to apply the proposed framework on more complex robotic systems.
Another interesting direction will be to explore how BO compares with other active sampling methods, e.g., CMA-ES, Simulated Annealing, etc.
Performing a thorough theoretical analysis of the proposed framework is another promising direction.
Finally, it will be interesting to use the computed distance bound for the safety analysis and verification of synthesized controllers.

\section*{Acknowledgments} \label{sec:acknowledge}
This research is supported by NSF under the CPS Frontiers VehiCal project (1545126), by the UC-Philippine-California Advanced Research Institute under project IIID-2016-005, by the ONR MURI Embedded Humans (N00014-16-1-2206), by the Army Research Laboratory \footnote{The views and conclusions contained in this document are those of the authors and should not be interpreted as representing the oﬃcial policies, either expressed or implied, of the Army Research Laboratory or the U.S. Government. The U.S. Government is authorized to reproduce and distribute reprints for Government purposes notwithstanding any copyright notation here on.} and was accomplished under Cooperative Agreement Number W911NF-17-2-0196; and in part by Toyota under the iCyPhy center.

\section*{Appendix}
\subsection{Proof of Proposition~\ref{prop:prob_gua}}
We denote by $R_n = \sum_{i=1}^{n} \dist^* - \dist^{BO}(i)$, the cumulative regret from $n$ iterations of BO, where $\dist^{BO}(i)$ is the $i$-th sample returned by BO. 

If $\dist \sim GP(0, k(\param, \param^{'}))$ with $k(\param, \param^{'})$ being a Mat\'ern kernel, then it can be shown that (see Theorem 2 in \cite{srinivas2009gaussian}) for any $\epsilon \in (0,1)$
\begin{equation}
    \text{Pr}(R_n \leq \mathcal{O}(\sqrt{|\paramspace|\cdot n \gamma_n}) \quad \forall n \geq 1) \geq 1 - \epsilon
    \label{eqn:Thm2}
\end{equation}
where $|\paramspace|$ is the dimension of the $\paramspace$, $n$ is the number of iterations of BO and $\gamma_n$ is the \textit{maximum information gain} after $n$ rounds.
$\gamma_n$ for a Mat\'ern32 kernel is given by (see Theorem 5 in~\cite{srinivas2009gaussian}), 
\begin{equation*}
    \gamma_n = \mathcal{O}(n^{\frac{|\paramspace|(|\paramspace|+1)}{(3+|\paramspace|(|\paramspace|+1))}}(\log n )).
\end{equation*}

We know that, $\hat{\dist}^{BO}(n) \geq \dist^{BO}(i), \forall i \in \{1, \dots, n\}$ since $\hat{\dist}^{BO}(n)$ is the maximum among all the $n$ iterations of BO.
Hence, we have, $R_n = \sum_{i=1}^{n} \dist^* - \dist^{BO}(i) \geq \sum_{i=1}^{n} \dist^* - \hat{\dist}^{BO}(n) = n \cdot (\dist^* - \hat{\dist}^{BO}(n))$. 
\eqref{eqn:Thm2} can thus be written as, 
\begin{equation}
\begin{split}
    \text{Pr}(n \cdot (\dist^* - \hat{\dist}^{BO}(n)) \leq \mathcal{O}(\sqrt{|\paramspace|\cdot n \gamma_n})\; \forall n \geq 1) &\geq 1 - \epsilon \\
    \text{Pr}(\dist^* > \hat{\dist}^{BO}(n) + \mathcal{O}(\sqrt{|\paramspace| \cdot \gamma_n/ n})\; \forall n \geq 1) &< \epsilon
\end{split}
\end{equation}


Ignoring the polylog factors and using $\gamma_n$ for the Mat\'ern $3/2$ kernel, we have, 
\begin{equation}
    \text{Pr}(\dist^* > \hat{\dist}^{BO}(n) + r_n\; \forall n \geq 1) < \epsilon 
    \label{Eqn:prop_gua}
\end{equation}
where $r_n = C\bigg(\sqrt{\frac{|\paramspace|}{n^{\frac{3}{3+|\paramspace|(|\paramspace|+1)}}}}\bigg)$ for some constant $C$.

Note that the above proof can also be used to obtain the convergence rates for different kernels by substituting the appropriate $\gamma_n$. 
$\gamma_n$ for commonly used kernels have been derived in Theorem 5 in~\cite{srinivas2009gaussian}.

\subsection{Proof of Corollary~\ref{cor:prob_val}}
Let there exist a $n > 0$ such that $\hat{\dist}^{BO}(n) < \tau - r_n$ where $\tau$ is the validation threshold and $r_n$ is defined as above. Rewriting~\eqref{Eqn:prop_gua} as, 
\begin{equation}
    \text{Pr}(\dist^* < \hat{\dist}^{BO}(n) + r_n\; \forall n \geq 1) \geq 1 - \epsilon
\end{equation}
we have,
\begin{equation}
    \text{Pr}(\dist^* < \tau) \geq 1 - \epsilon
\end{equation}
and we can conclude that the model $\abs$ is \textit{validated} for system $\sys$.

\bibliographystyle{plainnat}
\bibliography{RSS2018}

\end{document}